\documentclass[11pt,twoside]{article}

\usepackage{graphicx} 
\usepackage{mathptmx} 
\usepackage{amsmath}
\usepackage{amssymb}   

\usepackage{subfigure}
\usepackage{psboxit}
\usepackage{epsfig}
\usepackage{times}
\usepackage{wrapfig}
\usepackage{enumerate}
\usepackage{lscape}
\usepackage{longtable}
\usepackage{moreverb}
\usepackage{latexsym}
\usepackage{rotating}
\usepackage{natbib}
\usepackage[dvips]{color}

\def\aap{A\&A}%
\def\aj{AJ}%
\def\apj{ApJ}%
\def\mnras{MNRAS}%
\def\pasp{PASP}%

\begin{document}

\markboth{C. Papadaki et al. }{IP Peg in outburst: Echelle spectroscopy \& Modulation Doppler Tomography}
\title{\parbox{\textwidth}{ {\normalsize \sl 2008: The Journal of Astronomical Data 14, 1.    \hfill
\copyright\ C. Papadaki et al. }}\\ 
{\vspace{5mm} IP Pegasi in outburst: Echelle spectroscopy \& Modulation Doppler Tomography\footnote{ESO Proposal 63.H-0010(A).}}
}

\author{C. Papadaki$^{1}$, H.M.J. Boffin$^{2}$ and D. Steeghs$^{3}$\\ \vspace{3mm}  \\
\small (1) Vrije Universiteit Brussel, Pleinlaan 2, 1050 Brussels, Belgium\\
\small (2) ESO, Karl-Schwarzschild-Str. 2, 85748 Garching, Germany\\
\small (3) Department of Physics, University of Warwick, Coventry CV4 7AL, UK
}

\date{\small Received  June 2008, accepted 2008}
\maketitle

\section*{Abstract}

We analyse a unique set of time-resolved echelle spectra of the dwarf nova IP Peg, obtained at ESO's NTT with EMMI. The dataset covers the wavelength range of 4000-7500$\AA$ and shows Balmer, HeI, HeII and heavier elements in emission. IP Peg was observed one day after the peak of an outburst.

The trailed spectra, spectrograms and Doppler maps show characteristics typical of IP Pegasi during the early stages of its outburst, such as prograde rotation from the accretion disc flow, chromospheric emission from the secondary and spiral arms. The high-ionisation line of HeII $\lambda$4686 is the most centrally located line and has the greatest radial extension compared to the HeI lines. The Balmer lines extend from close to the white dwarf up to $\approx0.45R_{\rm L}$, with the outer radius gradually increasing when moving from H$\delta$ to H$\alpha$. The application, for the first time, of the modulation Doppler tomography technique, maps any harmonically varying components present in the system configuration, and this variability information is not considered in standard Doppler tomography studies. We find, as expected, that part of the strong secondary star emission in Balmer and HeI lines is modulated predominantly with the cosine term, consistent with the emission originating from the irradiated front side of the mass-donor star, facing the accreting white dwarf. For the Balmer lines the level of the modulation, compared to the average emission, decreases when moving to higher series. Emission from the extended accretion disk appears to be only weakly modulated, with amplitudes of at most a few \% of the non-varying disk emission. We find no evidence of modulated emission in the spiral arms, which if present, is relatively weak at that our signal-to-noise ratio was good enough to put a lower detection limit of any modulated emission at 5--6\%. Only in one arm of the HeII $\lambda$4686 line, is there a possibility of modulated emission, but again, we cannot be sure this is not caused by blending with the nearby Bowen complex of lines.

 
      \section{Introduction}

      Cataclysmic variable stars (CVs) consist of a Roche-lobe filling main sequence star, called the secondary star, orbiting a white dwarf - the primary star. Mass flows from the secondary star through the inner Lagrangian point towards the primary star, ultimately leading to the formation of an accretion disc (AD). After the AD's formation, a bright spot (BS) is formed at the contact point where the gas stream hits the AD. Dwarf novae (DNe) are a subclass of cataclysmic variables with two different brightness states, quiescence and outburst. In the latter, the system usually brightens by 2--5\,mag with a time-scale that can vary from tens to hundreds of days. DNe can also show a third brightness state, the superoutburst, where the system brightens by typically an additional magnitude or two compared to normal outburst.

      IP Peg is the brightest eclipsing ($i \approx81^{\circ}$) DN above the period gap, and was discovered by \citet{lip81}. It has a 3.8-h orbital period and a $V$ magnitude in the range of 14--12\,mag, the extreme values corresponding to quiescence and outburst, respectively. 

      From the first photometric light curves during quiescence and return to quiescence  \citep{woo86,woo89}, it was evident that the BS produces a prominent orbital hump, when it rotates into view from behind the disc with a single-stepped ingress and double-stepped egress superimposed on a more gradual disc-eclipse. Infrared light curves showed 0.2\,mag ellipsoidal variations from the secondary star superposed on a deep eclipse of the WD and the BS \citep{szk86}. \citet{woo89}, based on the analysis of 39 eclipses, found out that IP Peg has a rapidly increasing orbital period. However, a new study performed by \citet{wol93} and combining 49 eclipses along with previously reported ones,  reached the conclusion that the orbital period varies sinusoidally with a period of 4.7 years. This variation they attributed to a third body, a late M dwarf star with a most probable mass of 0.10\,$\rm M_\odot$. They also showed that the rate of shrinkage in disc radius during decline from outburst is constant during different outburst cycles, following an exponential decay. A study of the flickering in IP Peg \citep{bru00} showed that the BS is the dominant source of flickering, with the AD and WD contributing practically nothing. 

Spectroscopy of the red star \citep{mar89} showed TiO bands and the NaI doublet (red dwarf features), several broad double-peaked emission lines (disc features) and a narrow third peak in the emission lines moving in anti-phase with the double peaks (a chromospheric emission feature). \citet{bee00} performed a TiO study of IP Peg and concluded that TiO instead of the NaI doublet should be used for radial velocity studies of the secondary star. In this way they found that the contamination from disc emission, which in other studies returned an elliptical orbit, could be avoided and higher signal-to-noise ratio could be achieved. They gave a final projected orbital velocity of the companion star $K_2=331.3\,\rm km\,s^{-1}$ and refined values of the primary and secondary star masses $M_1=1.05\,\rm M_\odot$ and $M_2=0.33\,\rm M_\odot$, respectively. Covering a complete cycle from quiescence to outburst and quiescence again, the spectral characteristics of IP Peg will now be described. 

{\it Quiescence spectra} \citep{mar88} show broad double-peaked emission lines on a blue continuum, a strong orbital hump in the continuum (but much weaker in the lines) and a flat Balmer decrement indicative of optically thick emission. The trailed spectra show the signature of a prograde rotation of the AD, where the blue-shifted emission is eclipsed prior to the red-shifted one, a weak S-wave component in the red-shifted wing and a 10\% blue/red asymmetry before eclipse, most probably arising from line emission from the BS. 

{\it Early rise-to-outburst spectra} \citep{wol98} show Balmer emission, TiO absorption bands, KI and NaI absorption doublets, a weak S-wave and two bright spots in both Doppler and eclipse maps. The study of \citet{wol98}, as well as a re-analysis of their results using an extension of the classical eclipse mapping method \citep{bob99}, concluded that the first spot should be identified as the BS while two possibilities existed for the second spot. It either was the beginning of formation of a spiral arm or a second inner BS due to gas-stream overflow as predicted by, e.g., simulations performed by \citet{arm96}.

{\it Rise-to-outburst spectra} \citep{har94} show a decrease in the equivalent width of H$\alpha$ emission line, down to a level similar to that observed immediately after outburst and, simultaneously, weak H$\alpha$ emission from the secondary due to irradiation of the secondary star by the boundary layer.

{\it Near-outburst spectra} \citep{pic89} show unusually strong high-excitation lines, Balmer lines in emission, indications of a very disturbed AD with a non-uniform He emission distribution, an outflowing wind from the WD vicinity and chromospheric emission from the secondary.    

{\it Outburst spectra} as presented by \citet{mar90} show: i) strong Balmer and HeII $\lambda$4686 double-peaked emission lines rising from the AD, ii) non-Keplerian emission filling in the centre of HeII $\lambda$4686, probably due to a compact outflowing wind or inflowing magnetic accretion columns from close to the WD, iii) Balmer but not HeII emission from the secondary, located in the polar region on the side facing the AD, iv) no evidence for an enhanced gas-stream compared to quiescence, v) large azimuthal asymmetry in the outer disc, with the blue-shifted side of the disc being brighter and possibly more extended at phase 0 and vi) a sharp transient component in the HeII line possibly generated from the interaction of the gas stream and disc. \citet{mar90} suggested that He emission can be explained as photoionisation of the disc by the boundary layer and found that Balmer emission in outburst can be attributed by 3/4 to the disc, 1/4 to irradiation of the red star and also to boundary layer irradiation. Another study performed by \citet{ste96} showed, from trailed spectra and Doppler maps, asymmetric disc emission, strong secondary star emission and a low velocity emission in the H$\alpha$ line only. This stationary emission, appearing as an emission blob in the centre of the Doppler maps, was interpreted as slingshot prominences, material trapped in magnetic loops and co-rotating with the secondary star.  During another outburst maximum, \citet{har99a}  found metal line emission such as MgII. Doppler tomography of the He and metal lines locates those lines' emission on the inner Roche lobe of the secondary star.

{\it Early decline-from-outburst spectra} \citep{hes89} reveal for the first time in such strength a blend of high-excitation lines of HeII, CIII and NIII along with a very strong chromospheric emission-line component from the secondary. 

{\it Decline-from-outburst spectra} \citep{mor00} show irradiation-induced emission from the companion star in Balmer, He, MgII and CII lines. This emission, located only near the poles of the secondary, suggests shielding of the equatorial regions of the secondary star by the vertically extended AD.

{\it Early-quiescence spectra} in the infrared \citep{lit01} reveal the existence of mirror eclipses in IP Peg. A mirror eclipse is an eclipse of the secondary star by an optically thin AD and appears in the trailed spectra as a reduction in the equivalent widths of the lines, at phase 0.5. A mirror eclipse also has a mirror-symmetry with the classical rotational disturbance, i.e. the red-shifted part of the line is eclipsed prior to the blue-shifted one.

Even though IP Peg had shown many peculiarities and novelties in the studies cited previously, it had to reveal another breakthrough. \citet{ste97} were the first to find convincing evidence for the existence of spiral structure in the AD of a close binary. Spiral waves had long been predicted in studies of tidal interactions and numerical simulations (see \citet{bof01} and references therein) and may play a crucial role in our understanding of the angular momentum transport. \citet{ste97} obtained spectrophotometric observations of IP Peg during the rise to an outburst, covering H$\alpha$ and HeI $\lambda$6678. Disc emission was centred on the WD and had a strong azimuthal asymmetry in the forms of two spiral arms. The repeatability of the two-armed spiral structure was confirmed during another outburst maximum. \citet{har99a} applied Doppler tomography of the HeII $\lambda$4686 line and clarified the shock nature of the spiral arms which showed a jump of a factor of more than 2 in intensity. Additionally they calculated an azimuthal extent of 90$^{\circ}$, a 30$^{\circ}$ upper limit in opening angle and a 15\% contribution of the shocks to the total disc emission. \citet{mor00} showed that the spiral pattern persists even 5 and 6 days after the outburst maximum, without showing any diminution in strength. Time-resolved optical spectra obtained during late outburst by \citet{bap00} reveal, through the eclipse mapping technique, that spirals are present even 8 days after the onset of an outburst. Their study showed a similar contribution to the continuum emission, as that measured during the peak of the outburst but also a retrograde azimuthal rotation of 58$^{\circ}$ compared to outburst maximum.  

\citet{bap05} performed a re-analysis of previous spectroscopic data of IP Peg during outburst \citep{mor00}, based on the eclipse mapping technique. The difference in their new study was based on the fact that they separated the inner and outer regions of the emission line light curves, obtaining in this way the ``core'' and ``spiral arms'' lights curves, respectively. In this way they improved the previously noisy view of the spiral arms and were able to trace them in detail. The two-armed spiral structures were clearly visible in all their eclipse maps, albeit the ``blue'' arm appearing farther out from the disc's centre than the ``red'' one. However, their crucial findings were the sub-Keplerian velocities along the spiral structures, as well as the clear correlation between the opening angle of the spirals and the outburst stage (the later in the outburst, the smaller the opening angle of the spirals). Both these findings greatly favour the interpretation of the spiral asymmetric structures as tidally-induced spiral shocks instead of mere radiation patterns on an otherwise unperturbed AD.
 
A study performed by \citet{ste01} presents a review of the spiral structure that far, and plots in detail the azimuthal angles and extent of the spirals for the HeII $\lambda$4686 emission line during an outburst presented by \citet{har99}. In more detail, it is shown that the two spirals can be traced for almost 180$^{\circ}$, while their velocity varies from 495 to 780\,km\,s$^{-1}$. Under the assumption of Keplerian velocities, they show that the arms cover a substantial part of the disc, 0.3--0.9 times the distance to the inner Langragian point $L_1$. It is pointed out that tracing the spiral arms and plotting their position and intensity as a function of azimuth throughout an outburst, or even better applying this procedure to a range of emission lines simultaneously, can yield important results on the evolution of the spirals. A future paper, as stated in the Discussion section, is intended to address this issue. A first approach on tracing the spiral arms of two different emission lines (H$\alpha$, HeI $\lambda$6678), as well as plotting their emissivity as a function of azimuth, is presented in \citet{ste01}. It is shown that, even though there is a similar position for the spirals in terms of velocity as a function of azimuth, the intensity modulation is significantly different.

      \begin{figure}
	\centering
	\includegraphics[width=13cm]{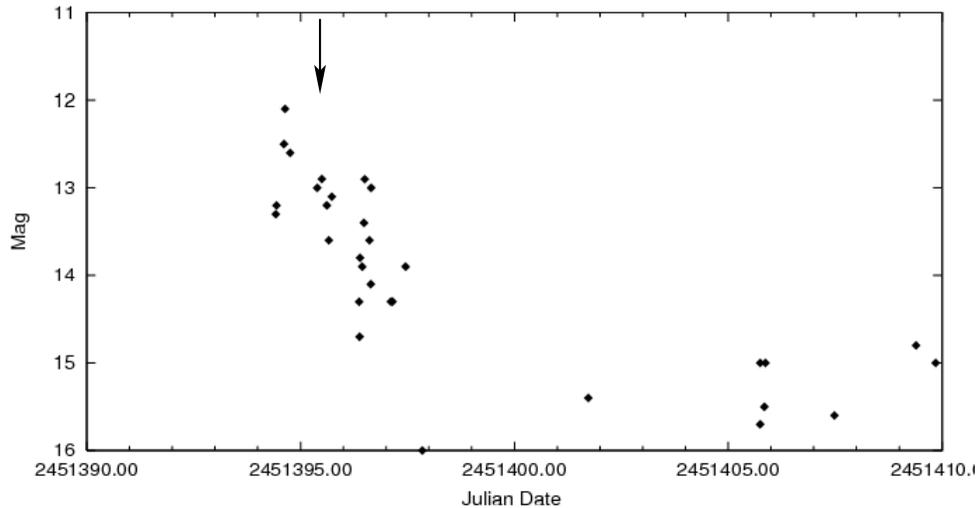}
	\caption{AAVSO long-term light curve of visual validated data. The arrow shows the timing of our observations, just 1 day after the peak of the outburst.}
	\label{f_aavso}
      \end{figure}

Our work is part of a project of studying CVs through high resolution time-resolved echelle spectra and applying the indirect imaging technique of Doppler tomography. The simultaneous study of several emission lines can yield important results on the AD structure, as presented for another CV system in \citet{pap08}. Following the introduction on IP Peg one can conclude on the benefits of applying such an analysis to this object, especially since it was observed during outburst. In this way we can study the behaviour of the spiral arms and map the structure of the AD in detail. The application of Modulation Doppler tomography, an extension of the classic Doppler tomography method \citep{ste03}, gives us the opportunity to map emission sources which can vary harmonically with the orbital period. Such an analysis is a first for IP Peg.

This paper is organised as follows: in Sect.~2 we present our observations, in Sec.~3 the data reduction, in Sect.~4 we perform the spectral analysis and Sect.~5 deals with the application and results of Modulation Doppler tomography. Last, Sect.~6 is devoted to the discussion and summary of our results.    

      \section{Observations}

      The observing run was conducted on the $4^{th}$ of August 1999, at ESO's NTT. Echelle spectroscopy was performed using the REMD mode of the EMMI spectrograph \citep{dek86}. The echelle grating \#9 and cross-disperser \#3 resulted in 21 orders covering the wavelength range 4000--7500\,\AA. The dispersion increased between 0.09 and 0.17\,\AA\,pix$^{-1}$, from blue to red, respectively. IP Peg was observed for 5.7\,h, covering 1.5 binary orbits. In total, 41 spectra with an exposure time of 300\,sec each, were obtained. 
 
Fig.~\ref{f_aavso} shows a long-term light curve of our system as generated from the AAVSO light curve generator tool\footnotemark[1]\footnotetext[1]{ http://www.aavso.org/data/lcg/}, choosing only visual-validated data to appear, and a window of a few days around outburst. An inspection of the outburst maximum reveals that IP Peg was at the peak of its outburst just 1 day before our observing run, when it was also 1 magnitude brighter.

\section{Data reduction}

      The resulting spectra were reduced using the standard echelle reduction procedures in {\it IRAF}\footnote{{\it IRAF} is distributed by the National Optical Astronomy Observatory, which is operated by the Association of Universities for Research in Astronomy, Inc., under cooperative agreement with the NSF.} and the slit spectra reduction task {\tt doecslit}. Whenever, both an interactive and non-interactive mode existed for a task, the interactive one was always chosen.

      Initially, the bias and lampflat frames were combined into one bias and flat frame with the {\tt zerocombine} and {\tt flatcombine} tasks, respectively. The spectra were then de-biased and a normalised flat was produced with the task {\tt apflatten} in order to apply the flat-fielding process. {\tt Apflatten}, after finding and tracing all apertures (defined as the sum of dispersion lines containing the signal of an order), also models both the profile and overall spectrum shape and removes it from the flat-field leaving only the sensitivity variations. The resulting normalised flat, if plotted perpendicular to the dispersion gives alternating smooth and noisy sections, smooth sections corresponding to the regions between the orders and having a value of 1, and noisy sections corresponding to the regions  inside the orders and having values fluctuating around 1. The spectra were then divided by the normalised flat and thus flat-fielded. At this stage, the spectra were also corrected for cosmic rays with the use of the {\tt cosmicrays} task, part of the {\tt crutil} package.

      The task {\tt doecslit} connects and combines several individual tasks to provide a single, complete data reduction path. It provides a degree of guidance, automation and record keeping. Additionally, the user with the choice of the interactive mode (used in this set of spectra) and the use of a graphical interface, is able to be in control of the whole procedure, and if necessary, fine-tune the reduction process while it runs. The several steps involved in {\tt doecslit} and therefore applied in the dataset are summarised as follows:

\begin{itemize}

\item In an aperture reference image, preferably a bright object, the orders were located and given a default fixed aperture width, which could be changed interactively if needed. 

\item The background sample regions were inspected. The function fitting parameters could be set interactively in order to best subtract the background later on.

\item Apertures were then defined for all object and standard star images, using as initial apertures, those defined from the aperture reference image. All images sequentially passed through this step. Possible deviations between the aperture reference image and the rest of the images, were fixed interactively by re-tracing and re-centering the apertures of the images.

 \item The actual extraction of the spectra was done using the {\it variance} option in the {\it weights} parameter of the {\tt sparams} task of the {\tt echelle} package. Variance weighting, often called optimal extraction, is performed by summing the pixel values across each aperture at each point along the dispersion. Each pixel in the sum is weighted by its estimated variance based on a spectrum model and the detector readout noise and gain parameters. For detailed information on the {\it optimal} extraction method the reader should refer to \citet{hor86} and \citet{marsh89}, whose work was followed by {\it IRAF}.

\item Dispersion correction was then applied through the use of Thorium-Argon arc lamp spectra. Initially, a few arc lines were identified in each order (beginning, end and middle of the dispersion axis), making use of the Th-Ar line atlas provided by ESO for the specific {\it EMMI} setup used in the observing run. Additional lines were then automatically identified with the use of an identification Th-Ar line list available in {\it IRAF}. A dispersion function was then fit to the total of identified lines in each order, and then applied to the corresponding orders of the rest of the images. The typical root mean square of the fits was $\approx$0.012\,\AA.

\item For the flux calibration we used spectra of the spectrophotometric standard LTT 7379. First, the bandpass regions were interactively defined and then the standard calibration data of LTT 7379, available in {\it IRAF}, were interpolated to the user-defined bandpasses. The flux values for each aperture were then fit with a smooth curve, in order to relate system sensitivity to wavelength for each aperture, thus resulting in a sensitivity curve for every aperture.

\item Finally the standard star spectra were extinction corrected  (using the extinction file provided by ESO for the location of the La Silla Observatory), and flux calibrated using the derived sensitivity curves.

\end{itemize}
      
As a last step in {\tt doecslit}, all previous steps were applied to all object spectra, resulting in the final flux-calibrated spectra of IP Peg.


\begin{sidewaysfigure}[p]
 \centerline{\epsfig{figure=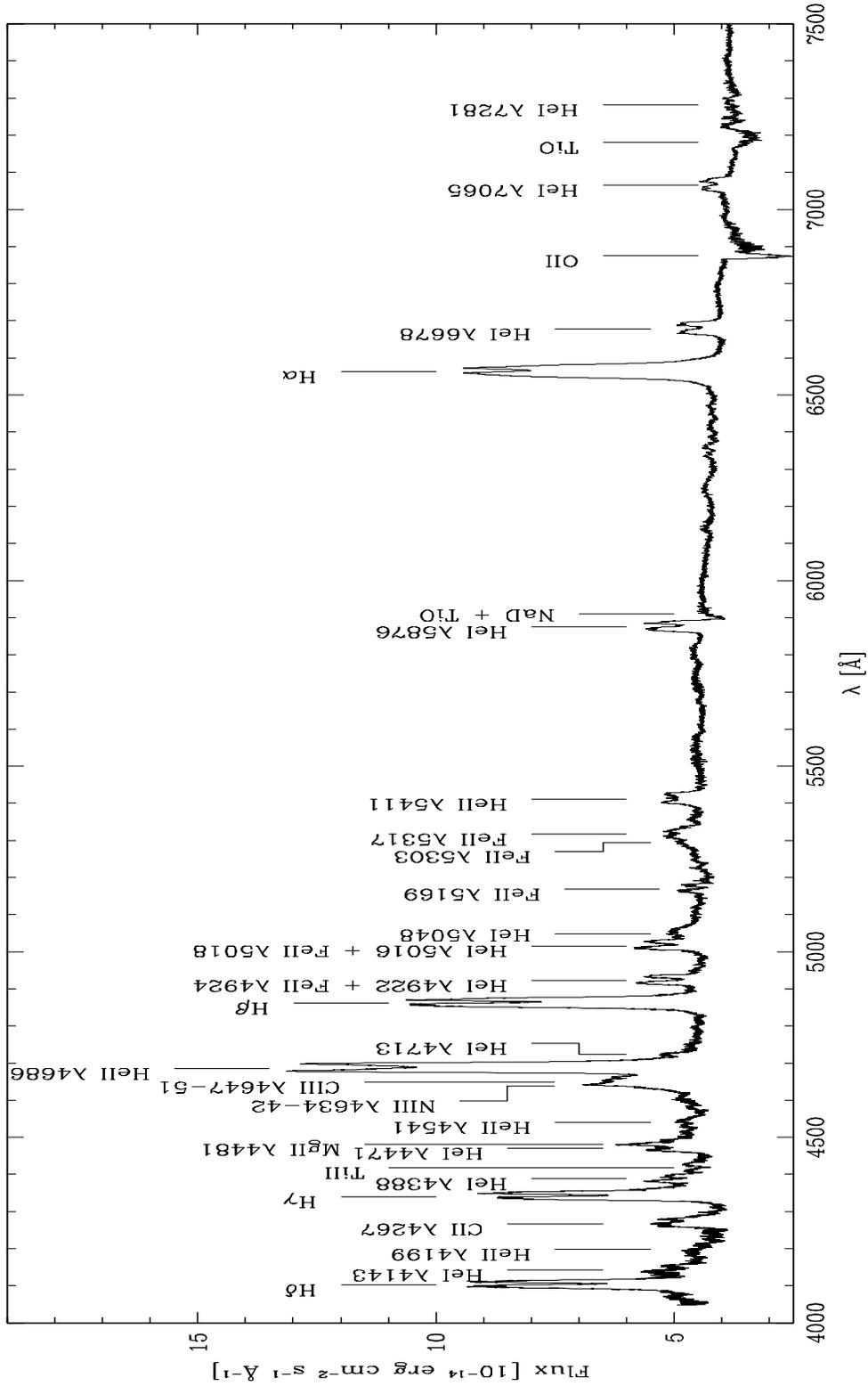,width=1\textheight,height=.65\textwidth}} 
 \caption{IP Peg average out-of-eclipse spectrum.}
 \label{f_avgsp_IPPeg}
 \end{sidewaysfigure}


\section{Spectral analysis}

The average out-of-eclipse (phases 0.15--0.85, where phase 0 is that of mid-eclipse) spectrum is shown in Fig.~\ref{f_avgsp_IPPeg}. To check that spectra of incomplete orbits do not affect the line profiles of the resulting average out-of-eclipse spectrum we first computed and compared the average out-of-eclipse spectrum of the whole dataset (1.5 orbits) and that of one full orbit. The spectrum resulting from the whole dataset differed only in the Balmer lines which showed an enhancement of the blue-peak intensity. The rest of the lines showed minimal or no differences at all. Therefore, the out-of-eclipse spectra of one full orbit were used in Fig.~\ref{f_avgsp_IPPeg}. The orbital phases of the individual spectra were calculated according to the linear orbital ephemeris given by \citet{wol93}:

\begin {equation}
  T_{0}[\rm HJD]=2445615.4224+0.^{\rm d}15820616 E
\end {equation}
where $T_{0}$ is the time of mid-eclipse.

\begin{table}
  \caption {Characteristics of emission lines.}
  \label{t_lin_ippeg}
  \centering
  \begin{tabular}{llll}
    \hline\hline
    & FWZI & EW & Integrated Flux\\
    & [km\,s$^{-1}$] & [\AA] & [$10^{-13}$ erg cm$^{-2}$ s$^{-1}$]\\
    \hline
    H$\delta$ & 5695 & 24.8 & 11.1\\
    H$\gamma$ & 5945 & 30.6 & 12.5\\
    HeII  $\lambda$4686 & 5055 & 51.3 & 23.8\\
    H$\beta$  & 4237 & 36.4 & 16.3\\
    HeI  $\lambda$4922+FeII $\lambda$4924 & 2101 & 4.7 & 2.2\\
    HeI  $\lambda$5876 & 1822 & 5.5 & 2.4\\
    H$\alpha$ & 4023 & 45.2 & 18.8\\
    HeI  $\lambda$6678 & 2479 & 6.7 & 2.7\\
    HeI  $\lambda$7065 & 1818 & 3.5 & 1.4\\
    \hline
  \end{tabular}
\end {table}

      \begin{figure}
	\centering
	\includegraphics[width=13.5cm]{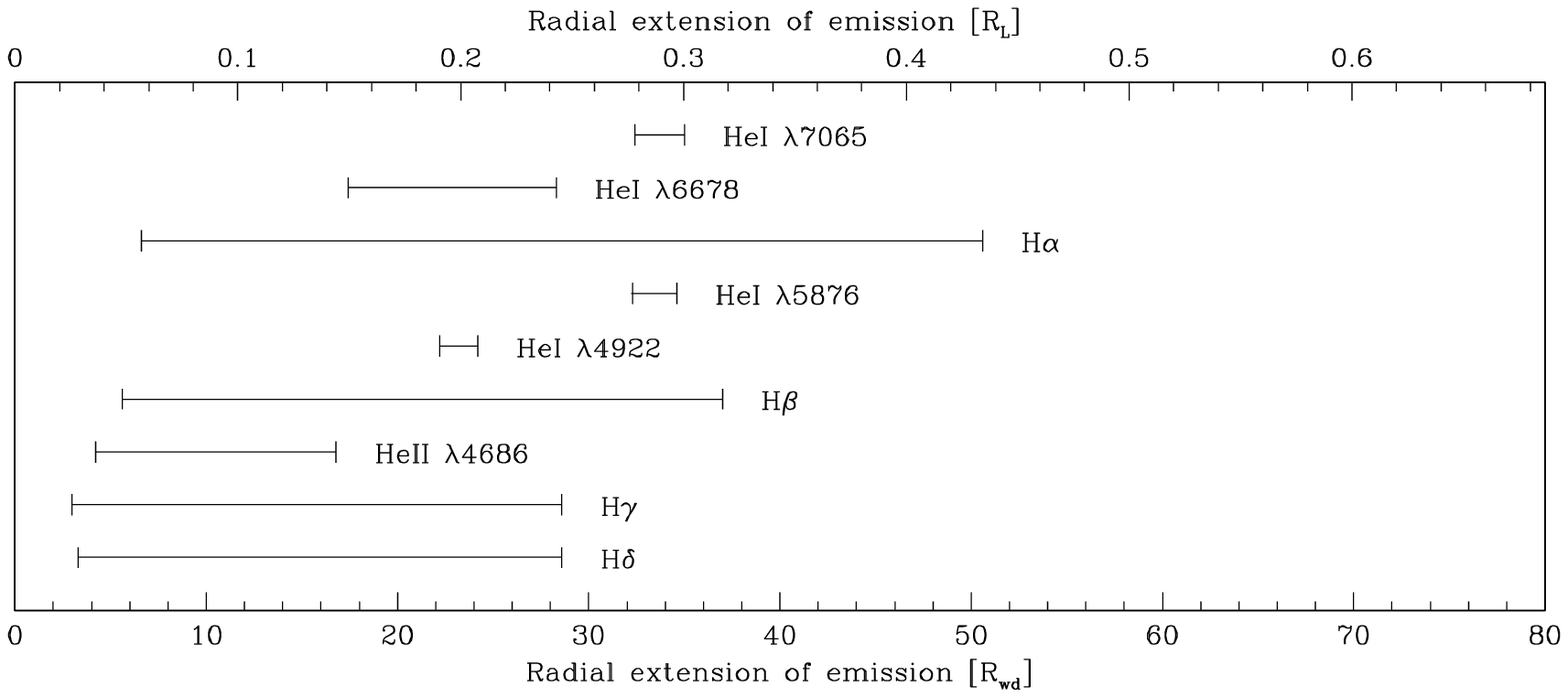}
	\caption{Radial extension of all emission lines. Bottom x-axis gives the radial extension as a factor of $R_{\rm wd}$, while top x-axis gives it as a factor of the volume radius of the primary Roche lobe $R_{\rm L}$}
	\label{f_radii_ippeg}
      \end{figure}

The outburst spectrum (Fig.~\ref{f_avgsp_IPPeg}) is characterised by several double-peaked emission lines, along with absorption features. There is strong Balmer and high-excitation emission from HeII $\lambda4686$ and the CIII-NIII blend. Also present is weaker emission from HeI ($\lambda$4143, 4388, 4471, 4713, 4922, 5016, 5048, 5876, 6678, 7065 and 7281), HeII ($\lambda$4199, 4541 and 5411) and CII $\lambda$4267 lines. IP Peg also shows emission from the heavier elements of  FeII (FeII 42 triplet and FeII $\lambda$5169, $\lambda$5303), TiII $\lambda$4418 and MgII $\lambda$4481. Finally, there are the NaD doublet and TiO absorption bands. 

      \begin{figure*}
	\centering
	\includegraphics[width=13.5cm]{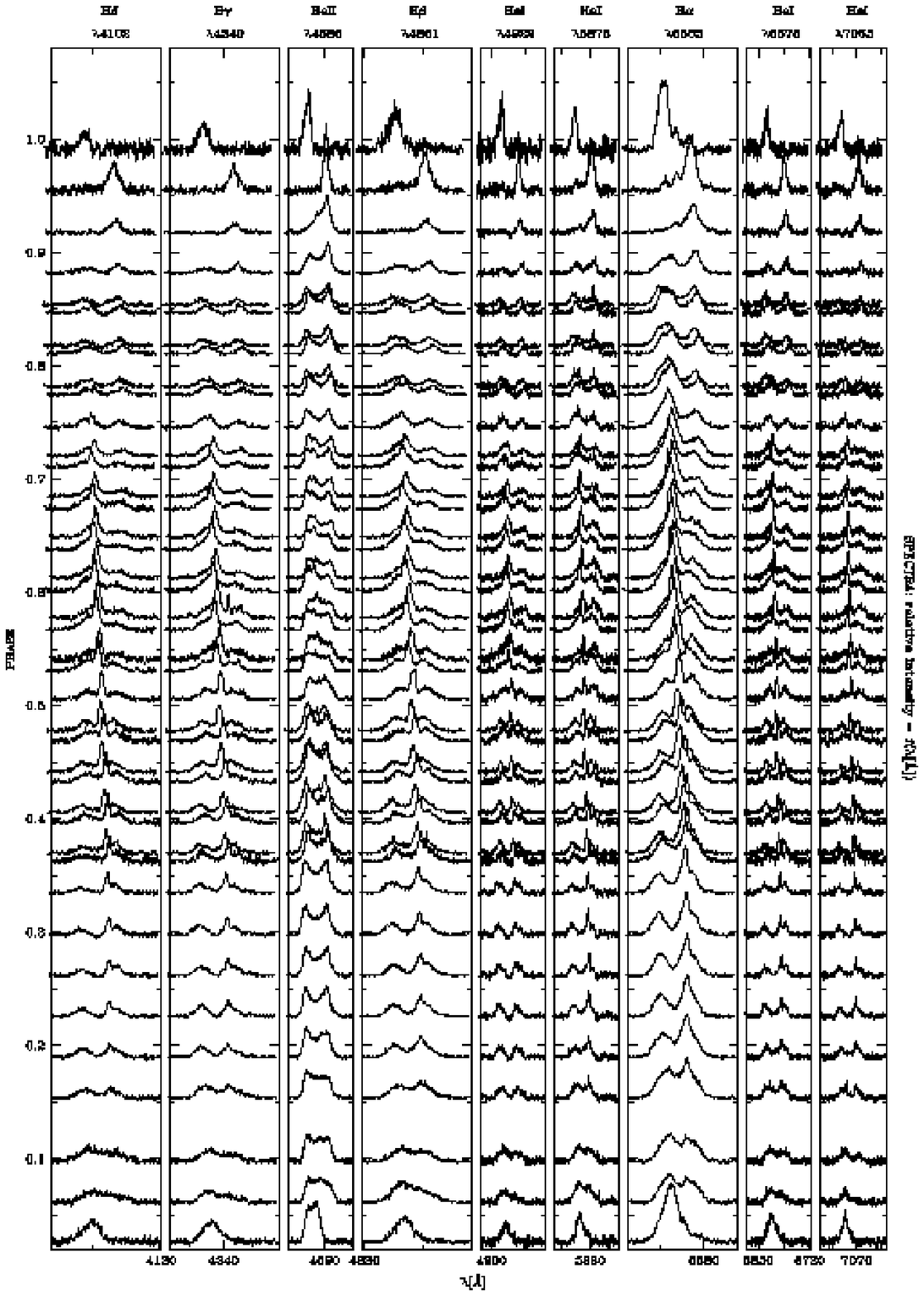}
	\caption{IP Peg trailed spectra of the 9 most prominent lines. The spectra have been ordered according to phase. The size of the x-axis small and big tickmarks are 35 and 70\AA, respectively.}
	\label{f_trsp_IPPeg}
      \end{figure*}

      \begin{figure*}
	\centering
	\includegraphics[width=13.5cm]{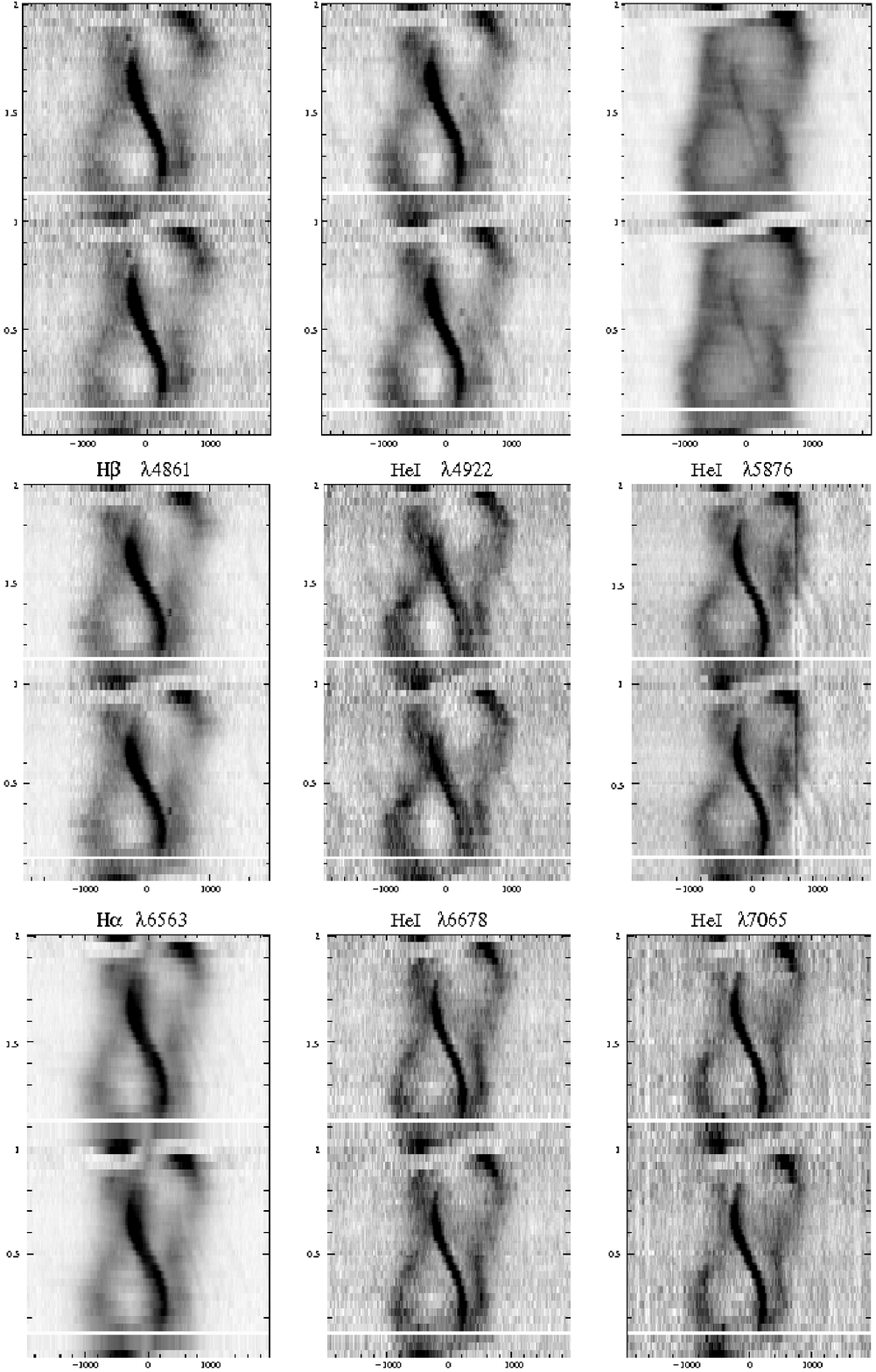}
	\caption{IP Peg spectrograms. The x-axis represents the velocity in km\,s$^{-1}$ and the y-axis the orbital phase. The grayscale contrast has been adjusted to emphasise between emission (black) and absorption (white).}
	\label{f_spec_IPPeg}
      \end{figure*}

All emission lines appear with symmetric line wings, except the lines of the blue part of the spectrum where one cannot tell with certainty due to their severe blending with neighbouring lines. Concerning the blue-to-red peak strength, the Balmer lines seem to differentiate from the high-excitation lines. They have an equal blue-to-red-peak strength (except H$\gamma$ which has a stronger red peak) while the high excitation lines (blended or stand-alone) appear to have a stronger blue peak. For the HeI lines it is harder to distinguish between the strength of the peaks, as they are either located in the wings of stronger lines ($\lambda$4143, $\lambda$4388, $\lambda$4713) or blended with equally strong lines ($\lambda$4471, $\lambda$4922, $\lambda$5016, $\lambda$4143). However, the stand-alone HeI lines ($\lambda$5876, $\lambda$6678, $\lambda$7065) point to a similar behaviour with the Balmer lines.

In this study, we will focus on the 9 most prominent emission lines i.e.~the Balmer lines, HeII $\lambda4686$ and HeI ($\lambda$4922, 5876, 6678, 7065). Table~\ref{t_lin_ippeg} shows the characteristics of the emission lines (full width at zero intensity -FWZI, equivalent width -EW, and integrated flux). However, caution should be drawn in the interpretation of these values for the severely blended lines, namely H$\delta$, H$\gamma$, HeII $\lambda4686$ and  HeI $\lambda5016$. The Balmer decrement from H$\alpha$ to H$\delta$ 1:0.77:0.68:0.55 shows a mix of optically thick and thin conditions.

Assuming a Keplerian velocity field and AD location of line emission, we can then estimate the radial extension of the emission lines given the fact that the velocities measured from the double-peak separation and FWZI are located in the outer and inner disc radius, respectively. Using equations (2) and (3) from \citet{pap08} and taking the tidal effect into account the resulting radial extension of the emission lines is shown in Fig.~\ref{f_radii_ippeg}. The bottom x-axis is measured in terms of the WD radius $R_{\rm wd}$, while the top x-axis in terms of the volume radius of the primary Roche lobe $R_{\rm L}$. The system parameters used to derive the inner and outer radius, $R_{\rm in}$ and $R_{\rm out}$, are the ones published by \citet{bee00}. The parameters $R_{\rm wd}$ and $R_{\rm L}$ were calculated according to equations (2.82) and (2.4c), respectively, in \citet{war03}. A look at Fig.~\ref{f_radii_ippeg} reveals a variety of behaviours. The Balmer line emission commences from close to the WD, $R_{\rm in}\approx 5\pm 2\,R_{\rm wd}$ and extends up to $R_{\rm out}\approx 50\,R_{\rm wd}$ for H$\alpha$. Both $R_{\rm in}$ and $R_{\rm out}$ gradually increase from H$\beta$ to H$\alpha$, albeit much more abruptly in $R_{out}$. The high-excitation line of HeII $\lambda$4686 also appears from close to the WD but is constrained to a smaller outer radius $R_{\rm out} \approx 17\,R_{\rm wd}$ and therefore is the most centrally located line.  The HeI $\lambda$7065, 5876 lines have a similar $R_{\rm in}\approx 32\,R_{\rm wd}$ and $R_{\rm out}\approx 35\,R_{\rm wd}$ but HeI $\lambda$4922 appears more centrally located, most probably due to its blending with the FeII line which also causes an asymmetry of the blue-to-red peak not present in the rest of the He lines as seen in Fig.~\ref{f_avgsp_IPPeg}. However, HeI $\lambda$6678 displays a different behaviour, being more centrally located and having a greater radial extension of $\approx 10\,R_{\rm wd}$, compared to the rest of the HeI lines that are more confined with an extension of $\approx 3\,R_{\rm wd}$. Non-uniform He distribution was also reported by \citet{pic89}, when IP Peg was observed near outburst maximum.

After normalising the spectra, and choosing the regions of interest, the trailed spectra of the 9 emission lines were computed. They are ordered according to orbital phase and shown in Fig.~\ref{f_trsp_IPPeg}. In order to enhance the emitting regions, the corresponding spectrograms are presented in Fig.~\ref{f_spec_IPPeg}, with black corresponding to maximum emission. They have been folded on phases 0--1 and repeated over two cycles. Both figures show characteristics typical of IP Peg: (i) prograde rotation of a disc-like emitting region, with the blue peak being eclipsed prior to the red one, (ii) double-peaked emission lines signature of an AD (ii) highly phase-dependent double-peaked separation, (iii) an S-wave moving in anti-phase with the peaks, signature of chromospheric emission from the secondary star. Additional features, seen around $\approx 1000\,\rm km\,\rm s^{-1}$ in the spectrograms of some emission lines, are caused by blendings with weaker emission lines in the red-wings of the principal lines. Each spectrogram has been plotted in a scale that best enhances its emitting regions, and therefore the intensities of the different emission lines should not be directly compared. At this point, the application of the modulation Doppler tomography, presented in the next section, will permit us to gain further insight of the AD structures.

\section {Modulation Doppler tomography}

\begin{sidewaysfigure}[p]
 \centerline{\epsfig{figure=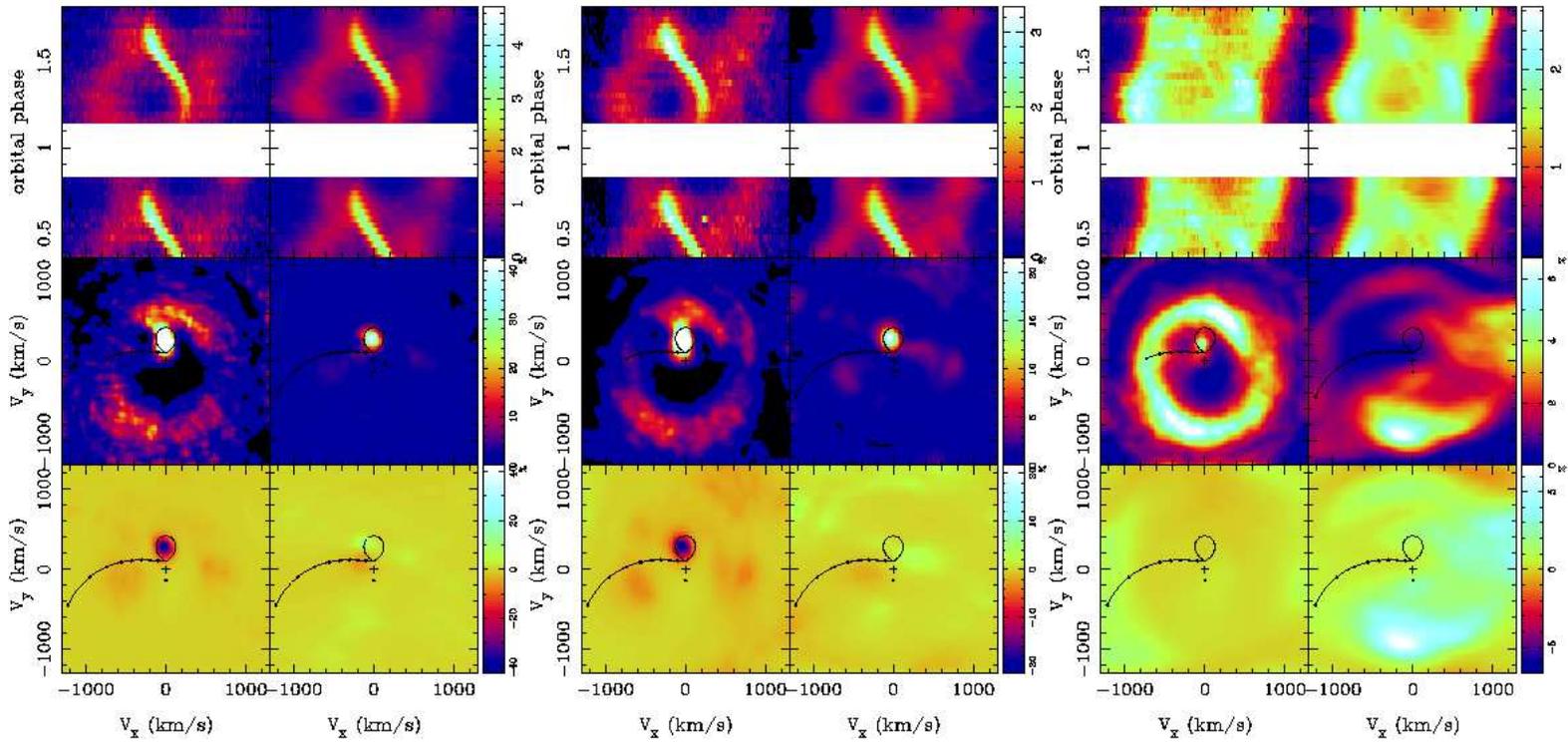,scale=0.7}} 
 \caption{In each panel from top to bottom and left to right: observed data, predicted data, non-variable emission distribution, total variable emission component,  cosine and sine terms of the variable component. From left to right: H$\delta$,  H$\gamma$, HeII $\lambda$4686.}
 \label{f_maps1}
 \end{sidewaysfigure}

\setcounter{figure}{5}

\begin{sidewaysfigure}[p]
 \centerline{\epsfig{figure=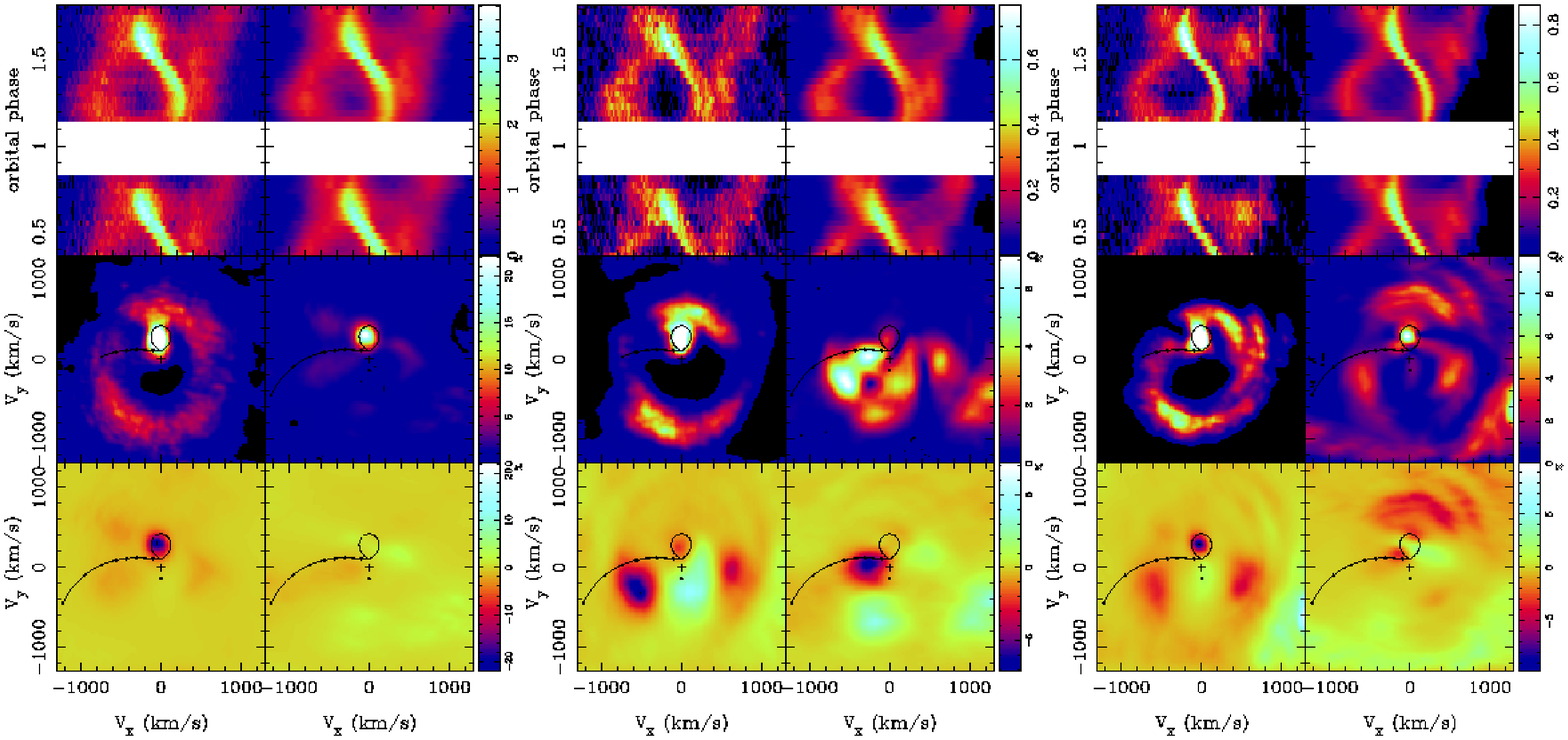,scale=0.7}} 
 \caption{(continued) From left to right: H$\beta$, HeI $\lambda$4922, HeI $\lambda$5876. }
 \label{f_maps2}
 \end{sidewaysfigure}

\setcounter{figure}{5}

\begin{sidewaysfigure}[p]
 \centerline{\epsfig{figure=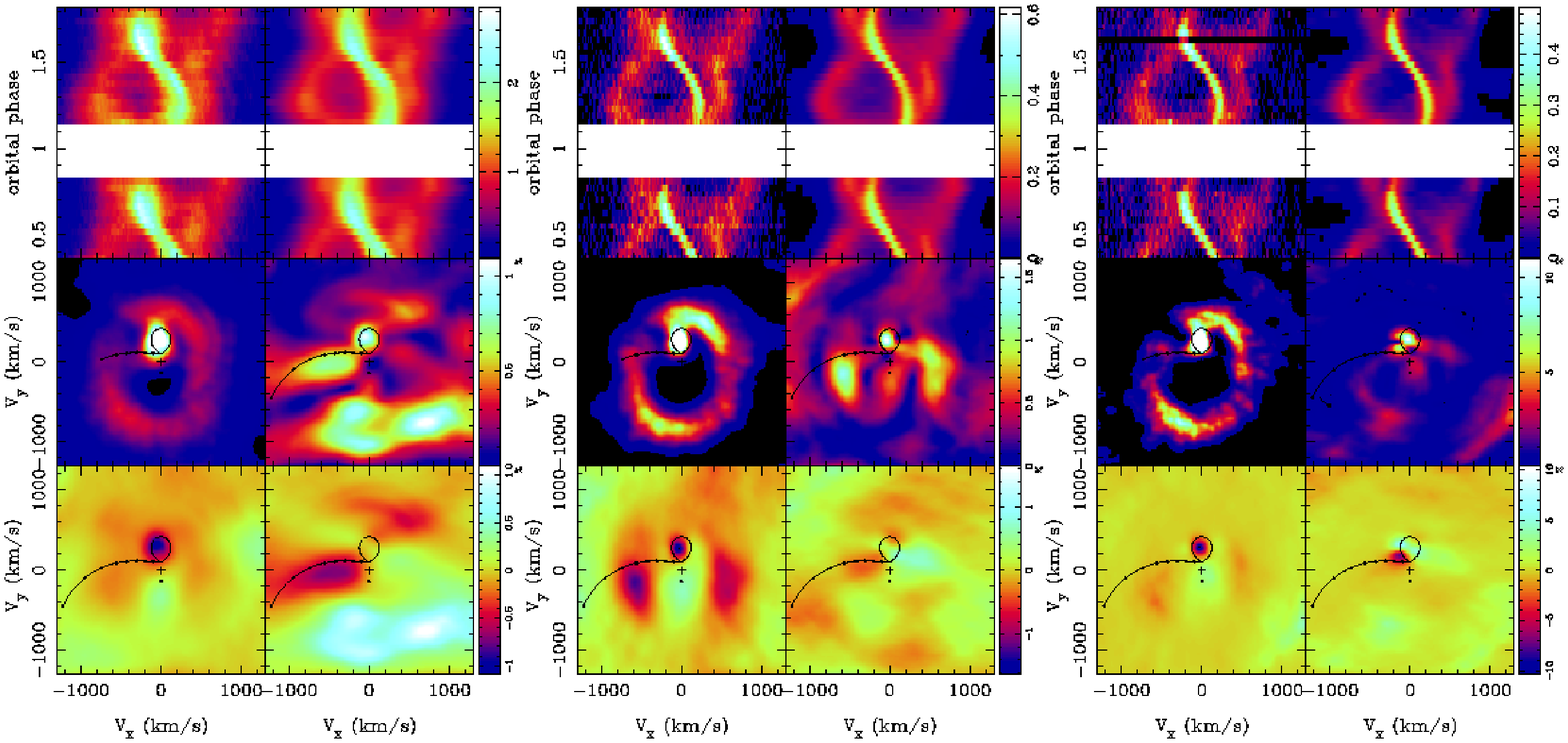,scale=0.7}} 
 \caption{(continued) From left to right: H$\alpha$, HeI $\lambda$6678, HeI $\lambda$7065.}
 \label{f_maps3}
 \end{sidewaysfigure}


Modulation Doppler tomography, developed by \citet{ste03}, is an extension of the classical Doppler tomography which relaxes one of its fundamental axioms, that the flux from any point is constant in time. This allows the mapping of time-dependent emission sources, and this variability information is not considered in standard Doppler tomography studies. Observations show that emission sources can modulate in time and this can happen either because there is anisotropic line-emission or because all sources are not equally visible at all times due to the system's geometry or both. Modulation Doppler tomography gives three output images; one describes the average non-varying flux distribution and the other two describe the sine and cosine term of the harmonically varying emission sources as a function of orbital phase. For a detailed description of Modulated Doppler tomography, the implementation of the MODMAP code and test reconstructions please refer to \citet{ste03} and references therein.

The resulting images after running MODMAP, for the 9 most prominent lines, are shown in Fig.~\ref{f_maps1}. From top to bottom and from left to right each panel shows: the observed and predicted data (the latter calculated from the constant and variable components represented by the maps), the non-variable emission distribution and the total variable emission component and finally the individual sine and cosine terms of the variable component. The input and predicted line profiles are plotted in the same scale, reflected by the colour wedge on their right. The average non-varying and total modulated emission maps are plotted in their own scale, with the colour wedge reflecting the scale of the modulated map. The scale is given as a percentage of the average emission to easily depict how strong the modulation is, relative to the constant component. The cosine and sine amplitude maps are also plotted in their own (common) scale, reflected by the colour wedge on their right, again as \% of the average emission, in order to ascertain their relative amplitudes. The Roche lobe shape and ballistic stream trajectory have also been computed by the MODMAP code, for the mass ratio and binary separation of IP Peg, and are over-plotted for reference.

The average non-varying-emission maps of all lines show the characteristic spirals plus donor star emission that is well known for IP Peg. The Balmer lines show strong modulation of the donor star, mostly attributed to its cosine term, while H$\alpha$ also shows modulated emission near the stream trajectory. The modulation of the donor star is at a 40\% level of the average emission in H$\delta$ and decreases when moving to higher Balmer series reaching about 1\% for H$\alpha$. There appears to be some modulated disc emission in H$\alpha$ though only at a 1--2\% level. No modulated flux, at the signal-to-noise ratio ($S/N$) of our data, is seen from the spiral arms.

The HeI lines reveal a strong component at the position of the secondary star which is of course modulated. Two disc regions appear to show some modulation, though only at the 1-2\% level. They also show a modulated component at the inner Lagrangian point, albeit of low level compared to average emission. The only exception is HeI $\lambda$4922 (see the cosine map) but one should note that this line is blended with FeII relatively strong in our set of spectra. Again, no significant amount of modulated flux from the spiral arms was needed to reproduce the observed data.

Finally, the high-ionisation line of HeII $\lambda$4686 also shows the characteristic spiral arms and secondary star emission like the rest of the lines and similarly to the maps Doppler maps presented by \citet{har99}.  The nearby Bowen blend causes some contamination, most likely the cause of the modulated structure. It is the only line in which there is a hint of modulated emission from the lower spiral arm at a 6\% level.

\section  {Summary \& Discussion}

This study presents the first time-resolved echelle spectra of IP Peg, one day after the peak of an outburst. Our spectra show the ensemble of the characteristics previously seen in spectroscopic studies of IP Peg, albeit not covering all lines together. The spectroscopic results, deduced from the average out-of-eclipse spectra are analytically given in Sec.~5 and can be summarised as follows:  
\begin{itemize}

\item Strong Balmer double-peaked emission arising from the AD. The lines show an equal blue-to-red-peak strength, as opposed to the higher excitation lines that have a stronger blue peak, therefore pointing to a less uniform emission. The Balmer decrement shows a mix of optically thin and thick conditions. The radial extension of the emission originates close to the white dwarf and extends up to $\approx 0.45 R_{\rm L}$ for H$\alpha$. The outer radius of the radial extension gradually increases when moving from H$\delta$ to H$\alpha$.

\item Strong double-peaked emission from the high excitation line of HeII $\lambda$4686 with a stronger blue peak. It is the most centrally located emission with its outer radius reaching $\approx 0.14 R_{\rm L}$.

\item Weaker HeI and heavier elements (FeII, TiII and MgII) double-peaked emission. Out of the four HeI lines ($\lambda$7065, 6678, 5876, 4922), HeI $\lambda$6678 is the most centrally located and shows the greater radial extension. 

\end{itemize}

The trailed spectra and spectrograms, reveal the prograde rotation, the double peaked emission lines arising from the disc, an S-wave component indicative of chromospheric emission from the secondary star and highly phase-dependent double-peaked separation, indicative of a distorted AD resulting in two asymmetric emission components, the spiral waves. All of these features, are present and obvious in the constructed Doppler maps.

We have applied to our data set, the modulation Doppler tomography method, which in addition to the classic Doppler maps reveals if any harmonically varying components are present in the system configuration. This is of great importance in sight of the spiral waves, which nevertheless have not shown any significant harmonically varying emission and therefore we can not claim anisotropic emission. The only line to show a hint of anisotropic emission, is the HeII line, and only in the lower spiral arm and at a 6\% level of the average emission, but this could be due to the blending of the line. However, the fact that for the majority of lines no modulated emission is seen from the spiral arms does not mean that the spirals do not modulate at all, but that no significant modulated contribution was needed to reproduce the observed data. At higher $S/N$ we may have seen some modulation. But, since we see features of amplitude a few \%, this means that our $S/N$ rules out strong anisotropy in disc emission. We could therefore conclude that, if present, the modulated emission of the spiral arms is relatively weak at that our $S/N$ was good enough to put a lower detection limit of any modulated emission at 5--6\%. 

The Balmer and HeI lines show strong secondary star emission, part of which is modulated. Its level of emission decreases when moving from H$\delta$ to H$\alpha$. As stated in the previous section secondary star emission is naturally modulated with the cosine term. This is normal since irradiation induced emission from the secondary is phase dependent, so it is natural to see a flux variation when the phase is changing. Given equation~3 of \citet{ste03} and that at phase 0.5--0.7, the face on secondary should emit more, then the cosine term is the dominant component.

All HeI lines show modulated emission in 2 regions of the AD, but are at a very low level. Only in $\lambda$4922 do they appear more pronounced. These AD modulated regions, seem to appear in the "gaps" between the spiral arms and modulate with the sine-term. This could be in accordance with the spiral arms shielding the disc areas just behind them and hiding a possible cosine-term of disc modulated emission as well. Moreover, the sine-term is expected considering their phasing is in between the spiral arms. Geometric shielding caused by the spirals was also proposed from \citet{ste03} and shielding of the secondary star by the AD was proposed by \citet{wat03}. Finally, most of the HeI lines show modulated emission near the inner Lagrangian point also covering a small region at the beginning of the stream trajectory. H$\alpha$ though, shows pronounced modulated emission for the bigger part of the stream trajectory.

In a future paper it is our aim to study the same dataset, but concerning the spiral arms of the different lines only. Tracing, providing the amplitude, and the contribution to the total flux of the spiral arms of all lines, as well  as finding the ratio of the brightness between the two spiral arms can yield valuable results for this unique set of time-resolved spectra.     

\section{The data}

\begin{itemize}
    
\item fig3.dat: an ascii file that gives the radial extension of the emission lines as a factor of both the white dwarf radius $R_{\rm wd}$ and the volume radius of the primary Roche lobe $R_{\rm L}$, as seen in Fig.~3. 
  
\item HJD\_phase.dat: an ascii file giving the heliocentric Julian date and the orbital phase of the 41 spectra.
  
\item spectra: a folder that contains,
  
  \begin{itemize} 
    
  \item the average out-of-eclipse spectrum of Fig.~2 as a gzipped fits file, named \\IPPeg\_avr.fits.gz and
    
  \item the normalised trailed spectra of the 9 most prominent emission lines as shown in Fig.~4. These trailed spectra were the ones used for the reproduction of Fig.~4--Fig.~6. The naming convention is \{emission line\}\_\{sequence number\}.fits.gz, resulting in 9 (number of emission lines)$\times$41 (number of spectra) different files. 
    
  \end{itemize}
  
\end{itemize}

\section*{Acknowledgments}
C.P. gratefully acknowledges a doctoral research  grant by the Belgian Federal Science Policy Office (Belspo).


\begin{thebibliography}{60}
\small
\expandafter\ifx\csname natexlab\endcsname\relax\def\natexlab#1{#1}\fi

\bibitem[{{Armitage} \& {Livio}(1996)}] {arm96} {Armitage}, P.~J. \& {Livio}, M. 1996, \apj, 470, 1024 \vspace{-2mm}

\bibitem[{{Baptista} {et~al.}(2000)}] {bap00} {Baptista}, R., {Harlaftis}, E.~T. \& {Steeghs}, D. 2000, \mnras, 314, 727 \vspace{-2mm}

\bibitem[{{Baptista} {et~al.}(2002)}] {bap02} {Baptista}, R., {Haswell}, C.~A. \& {Thomas}, G. 2002, \mnras, 334, 198 \vspace{-2mm}

\bibitem[{{Baptista} {et~al.}(2005)}] {bap05} {Baptista}, R., {Morales-Rueda}, L., {Harlaftis}, E.~T., {Marsh}, T.~R. \& {Steeghs}, D. 2005, \aap, 444, 201 \vspace{-2mm}

\bibitem[{{Beekman} {et~al.}(2000)}] {bee00} {Beekman}, G., {Somers}, M., {Naylor}, T. \& {Hellier}, C. 2000, \mnras, 318, 9 \vspace{-2mm}

\bibitem[{{Bobinger} {et~al.}(1999)}] {bob99} {Bobinger}, A., {Barwig}, H., {Fiedler}, H., {Mantel}, K.-H., {{\v S}imi{\'c} }, D. \& {Wolf}, S. 1999, \aap, 348, 145  \vspace{-2mm}

\bibitem[{{Boffin}(2001)}] {bof01} {Boffin}, H.~M.~J. 2001, in Lecture Notes in Physics, Berlin Springer Verlag, Vol.573, Astrotomography, Indirect Imaging Methods in Observational Astronomy, eds. {Boffin}, H.~M.~J. and {Steeghs}, D. and {Cuypers}, J., p. 69 \vspace{-2mm}

\bibitem[{{Bruch}(2000)}] {bru00} {Bruch}, A. 2000, \aap, 359, 998 \vspace{-2mm}

\bibitem[{{Dekker} {et~al.}(1986)}] {dek86} {Dekker}, H., {Delabre}, B. \& {Dodorico}, S. 1986, Presented at the Society of Photo-Optical Instrumentation Engineers (SPIE) Conference, Vol.627, Instrumentation in astronomy VI; Proceedings of the Meeting, Tucson, AZ, Mar. 4-8, 1986. Part 1 (A87-36376 15-35). Bellingham, WA, Society of Photo-Optical Instrumentation Engineers, 1986, p. 339-348 \vspace{-2mm}

\bibitem[{{Harlaftis}(1999)}] {har99} {Harlaftis}, E. 1999, \aap, 346, L73 \vspace{-2mm}

\bibitem[{{Harlaftis} {et~al.}(1994)}] {har94} {Harlaftis}, E.~T., {Marsh}, T.~R., {Dhillon}, V.~S. \& {Charles}, P.~A. 1994, \mnras, 267, 473 \vspace{-2mm}

\bibitem[{{Harlaftis} {et~al.}(1999)}] {har99a} {Harlaftis}, E.~T., {Steeghs}, D., {Horne}, K., {Mart{\'{\i}}n}, E. \& {Magazz{\'u}}, A. 1999, \mnras, 306, 348 \vspace{-2mm}

\bibitem[{{Hessman}(1989)}] {hes89} {Hessman}, F.~V. 1989, \aj, 98, 675 \vspace{-2mm}

\bibitem[{{Horne}(1986)}] {hor86} {Horne}, K. 1986, \pasp, 98, 609, \vspace{-2mm}

\bibitem[{{Lipovetsky} \& {Stepanian}(1981)}] {lip81} {Lipovetsky}, V.~A. \& {Stepanian}, J.~A. 1981, Astrofizika, 17, 573 \vspace{-2mm}

\bibitem[{{Littlefair} {et~al.}(2001)}] {lit01} {Littlefair}, S.~P., {Dhillon}, V.~S., {Marsh}, T.~R. \& {Harlaftis}, E.~T. 2001, \mnras, 327, 475 \vspace{-2mm}

\bibitem[{{Marsh}(1988)}] {mar88} {Marsh}, T.~R. 1988, \mnras, 231, 1117  \vspace{-2mm}

\bibitem[{{Marsh}(1989)}] {marsh89} {Marsh}, T.~R. 1989, \pasp, 101, 1032 \vspace{-2mm}

\bibitem[{{Marsh} \& {Horne}(1990)}] {mar90} {Marsh}, T.~R. \& {Horne}, K. 1990, \apj, 349, 593  \vspace{-2mm}

\bibitem[{{Martin} {et~al.}(1989)}] {mar89} {Martin}, J.~S., {Jones}, D.~H.~P., {Friend}, M.~T. \& {Smith}, R.~C. 1989, \mnras, 240, 519 \vspace{-2mm}

\bibitem[{{Morales-Rueda} {et~al.}(2000)}] {mor00} {Morales-Rueda}, L., {Marsh}, T.~R. \& {Billington}, I. 2000, \mnras, 313, 454 \vspace{-2mm}

\bibitem[{{Neustroev} {et~al.}(2002)}] {neu02} {Neustroev}, V.~V., {Borisov}, N.~V., {Barwig}, H., {Bobinger}, A., {Mantel}, K.~H., {{\v S}imi{\'c}}, D. \& {Wolf}, S. 2002, \aap, 393, 239 \vspace{-2mm}

\bibitem[{{Papadaki} {et~al.}(2008)}] {pap08} {Papadaki}, C., {Boffin}, H.~M.~J., {Steeghs}, D. \& {Schmidtobreick}, L. 2008, astro-ph/0804.0898, accepted by A\&A \vspace{-2mm}
\bibitem[{{Piche} \& {Szkody}(1989)}] {pic89} {Piche}, F. \& {Szkody}, P. 1989, \aj, 98, 2225 \vspace{-2mm}

\bibitem[{{Steeghs}(2001)}] {ste01} {Steeghs}, D. 2001, in Lecture Notes in Physics, Berlin Springer Verlag, Vol.573, Astrotomography, Indirect Imaging Methods in Observational Astronomy, eds. {Boffin}, H.~M.~J. and {Steeghs}, D. and {Cuypers}, J., p. 45 \vspace{-2mm}

\bibitem[{{Steeghs}(2003)}] {ste03} {Steeghs}, D. 2003, \mnras, 344, 448 \vspace{-2mm}

\bibitem[{{Steeghs} {et~al.}(1997)}] {ste97} {Steeghs}, D. and {Harlaftis}, E.~T. and {Horne}, K. 1997, \mnras, 290, 28 \vspace{-2mm}

\bibitem[{{Steeghs} {et~al.}(1996)}] {ste96} {Steeghs}, D., {Horne}, K., {Marsh}, T.~R. \& {Donati}, J.~F. 1996, \mnras, 281, 626 \vspace{-2mm}

\bibitem[{{Szkody} \& {Mateo}(1986)}] {szk86} {Szkody}, P. \& {Mateo}, M. 1986, \aj, 92, 483 \vspace{-2mm}

\bibitem[{{Warner}(2003)}] {war03} {Warner}, B. 2003, Cataclysmic Variable Stars (by Brian Warner, pp.~592.~ISBN 052154209X.~Cambridge, UK: Cambridge University Press, September 2003.) \vspace{-2mm}

\bibitem[{{Watson} {et~al.}(2003)}] {wat03} {Watson}, C.~A., {Dhillon}, V.~S., {Rutten}, R.~G.~M. \& {Schwope}, A.~D. 2003, \mnras, 341, 129 \vspace{-2mm}

\bibitem[{{Wolf} {et~al.}(1998)}] {wol98} {Wolf}, S., {Barwig}, H., {Bobinger}, A., {Mantel}, K.-H. \& {Simic}, D. 1998, \aap, 332, 984 \vspace{-2mm}

\bibitem[{{Wolf} {et~al.}(1993)}] {wol93} {Wolf}, S., {Mantel}, K.~H., {Horne}, K., {Barwig}, H., {Schoembs}, R. \& {Baernbantner}, O. 1993, \aap, 273, 160 \vspace{-2mm}

\bibitem[{{Wood} \& {Crawford}(1986)}] {woo86} {Wood}, J. \& {Crawford}, C.~S. 1986, \mnras, 222, 645 \vspace{-2mm}

\bibitem[{{Wood} {et~al.}(1989)}] {woo89} {Wood}, J.~H., {Marsh}, T.~R., {Robinson}, E.~L., {Stiening}, R.~F., {Horne}, K., {Stover}, R.~J., {Schoembs}, R., {Allen}, S.~L., {Bond}, H.~E., {Jones}, D.~H.~P., {Grauer}, A.~D. \& {Ciardullo}, R. 1989, \mnras, 239, 809  \vspace{-2mm}


\end{thebibliography}
\end {document}